\documentclass[aps,article,nofootinbib,groupedaddress]{revtex4}
\usepackage{graphicx}
\usepackage{amsmath,amssymb}
\usepackage{url,hyperref}
\usepackage[usenames]{color}

\hypersetup{
    colorlinks=true,
    linkcolor=red,
    citecolor=blue,
}

\def\be{\begin{equation}}
\def\ee{\end{equation}}
\def\ba{\begin{eqnarray}}
\def\ea{\end{eqnarray}}
\def\nn{\nonumber}

\def\f{\frac}

\newcommand\etal{{\em et al.}}
\newcommand\eg{{\em e.g.}}

\newcommand{\tc}{\textcolor{black}}
\begin{document}

\title{Testing gravity with CAMB and CosmoMC}

\date{\today}

\author{Alireza Hojjati$^{1}$, Levon Pogosian$^{1}$, Gong-Bo Zhao$^{2}$}

\affiliation{$^1$Department of Physics, Simon Fraser University, Burnaby, BC, V5A 1S6, Canada \\
$^2$Institute of Cosmology \& Gravitation, University of Portsmouth, Portsmouth, PO1 3FX, UK}

\begin{abstract}
We introduce a patch to the commonly used public codes CAMB and CosmoMC that allows the user to implement a general modification of the equations describing the growth of cosmological perturbations, while preserving the covariant conservation of the energy-momentum. This patch replaces the previously publicly released code MGCAMB, while also extending it in several ways. The new version removes the limitation of late-time-only modifications to the perturbed Einstein equations, and includes several parametrization introduced in the literature. To demonstrate the use of the patch, we obtain joint constraints on the neutrino mass and parameters of a scalar-tensor gravity model from CMB, SNe and ISW data as measured from the correlation of CMB with large scale structure.
\end{abstract}

\maketitle

\section{Introduction}

Future cosmological surveys will have the ability to measure the growth of large-scale structure with accuracy sufficient for discriminating between different models of dark energy (DE) and modified gravity (MG). To this aim, an interesting approach is to parametrize a general theory of gravity in a model independent way so that it reduces to general relativity (GR) for a choice of parameter values, and can also be related to certain classes of MG models. This approach has been taken by several groups to test the standard Lambda Cold Dark Matter ($\Lambda$CDM) model and to constrain parameters of various DE and MG models \cite{Linder:2007hg,Caldwell:2007cw,Hu:2007pj,Bertschinger:2008zb,Fang:2008sn,Zhao:2008bn,Giannantonio:2009gi,Zhao:2009fn,Daniel:2009kr,Daniel:2010ky,Song:2010rm,
Zhao:2010dz,Bean:2010zq,Song:2010fg,Lombriser:2010mp,Dossett:2011zp}. We refer the reader to \cite{Silvestri:2009hh,Clifton:2011jh} for a comprehensive review of the recent advances in the field of MG and ways of testing GR on cosmological scales.

Code for Anisotropies in the Microwave Background (CAMB) \cite{camb,Lewis:1999bs} uses the equations of GR together with Boltzmann (conservation) equations to calculate cosmological observables. It needs to be modified for studying the alternative gravity theories, as for example was done in \cite{Hu:2007pj,Fang:2008sn,Zhao:2008bn,Bean:2010zq}. One then needs a system of equations that is meaningful across a wide range of scales and redshifts. As stressed in \cite{Hu:2007pj}, the evolution of perturbations on super-horizon scales is governed by a set of consistency conditions which are separate from the sub-horizon dynamics. Namely, as shown in~\cite{Wands:2000dp,Bertschinger:2006aw}, in the absence of entropy perturbations, the space curvature defined on hyper-surfaces of uniform matter density, $\zeta$, must be conserved on scales outside the horizon in order to be consistent with the overall expansion of the universe. Hence, a consistent system of equations should decouple the super- and sub-horizon regimes. This separation of scales is made explicit in the parametrized Post-Friedmannian (PPF) Framework of~\cite{Hu:2007pj}, where a different systems of equations are used on super-horizon and sub-horizon scales. The advantage of the method advocated in this paper and used in MGCAMB (Modification of Growth with CAMB)~\cite{Zhao:2008bn} is that it employs a single system of equations across all linear scales, without sacrificing any of the important consistency conditions. The super-horizon and sub-horizon evolution decouple naturally, without having to be explicitly separated.

Here we introduce a generalized version of MGCAMB, which now has a form of a patch to CAMB and CosmoMC, available for download and public use at \cite{MGCAMB}. This version is capable of evolving all the species in the universe from early times to present. While the previous version was based on CAMBsources \cite{cambsources}, which made it difficult to use in CosmoMC, this version is based on CAMB and can be used easily with CosmoMC to constrain desired parameters. Several parametrizations from the literature are included in this version which makes it useful for investigating a wide range of models.

Massive neutrinos are not part of the standard $\Lambda$CDM model, and can be considered as an extension of the model that modifies the growth of large scale structure. It is, therefore, interesting to study possible degeneracies between effects of the neutrino mass and various MG or dark energy models which also modify the growth on large scales. We used MGCAMB to study the degeneracy between massive neutrinos and a specific class of $f(R)$ and Chameleon-type models using a combination of CMB,  supernovae and the CMB-LSS cross-correlation data of Ho et al. \cite{Ho:2008bz}. We find the degeneracy to be small due to the smallness of the overlap of the scales at which the respective modifications affect the observables we have considered.

This paper is organized as follows. In section \ref{perturbation}, we review the linear scalar perturbation equations for GR and introduce our modification to these equations. Some of the alternative parametrizations in the literature are also reviewed in this section. In Section \ref{MGCAMB}, we describe the implementation of our equations in CAMB and introduce the new MGCAMB patch. Joint constraints on massive neutrinos and some MG models are obtained in Section \ref{results}. Certain details of the way the modification was implemented in the CMB source function in CAMB and in the CMB-LSS cross-correlation patch of Ho et al. \cite{Ho:2008bz} are given in two appendices. We summarize in Section \ref{summary}.

\section{Formulation}
\label{perturbation}

As a starting point we assume that the background universe is described by a flat Friedmann-Robertson-Walker (FRW) metric, and consider linear perturbations in the metric and the energy-momentum tensor. In the Newtonian gauge, the perturbed line element has the form
\be  
\label{metric}
ds^2 = a(\tau)^2[-(1+2\Psi)d\tau^2+(1-2\Phi)dx^2] \,,
\ee
where $\tau$ is the conformal time and $\Psi$ and $\Phi$ are the two scalar metric potentials. In the synchronous gauge, which is used in CAMB, the perturbed line element is 
\be
\label{metric-sync}
ds^2 = a(\tau)^2[-d\tau^2+(\delta_{ij} + h_{ij})dx^i dx^j] \,,
\ee
where the scalar part of $h_{ij}$ is represented by functions $\eta$ and $h$ which in Fourier space are defined as \cite{Ma:1995ey}
\begin{equation}
h_{ij}(\vec{x},\tau) = \int d^3k e^{i\vec{k}\cdot\vec{x}} 	\left\{ \hat{k}_i\hat{k}_j h(\vec{k},\tau) +
(\hat{k}_i\hat{k}_j - {1 \over 3}\delta_{ij})\, 6\eta(\vec{k},\tau) \right\} \,,\quad \vec{k} = k\hat{k} \, .
\label{sync-metric}
\end{equation}
 We also consider perturbations in the radiation-matter content, which includes photons, neutrinos, baryons and cold dark matter (CDM):
\ba\label{en-mom-generic}
&&T^0_0+\delta T^0_0=-\rho(1+\delta)\, ,\nonumber\\
&&T^0_i+ \delta T^0_i =-(\rho+P)v_i\, ,\nonumber\\
&&T^i_j + \delta T^i_j=(P+\delta P)\delta^i_j+\pi^i_j\, ,
\ea
where $\delta\equiv \delta\rho/\rho$ is the density contrast, $v$ is the velocity field, $\delta P$ is the pressure perturbation and $\pi^i_j$ denotes the traceless component of the energy-momentum tensor perturbations. Unless noted otherwise, all the quantities are sums over all the species present in the universe. Working in Fourier space, it is convenient to introduce the comoving density perturbation $\Delta$:
\be
\rho \Delta = \rho \delta + 3 \frac{\mathcal{H}}{k}(\rho +P)v \ , 
\ee
where $\mathcal{H}\equiv\dot{a}/a$, and throughout this paper the overdot denotes the derivative with respect to the conformal time $\tau$. Note that we also use the physical Hubble parameter, $H\equiv\mathcal{H}/a$, later in the text. The anisotropic stress $\sigma$ and the momentum perturbation $\theta$ are defined as
\ba
&&(\rho+P)\sigma\equiv -(\hat{k}^i\hat{k}_j-\f{1}{3}\delta^i_j)\pi^i_j~,\\
&&(\rho+P) \theta \equiv ik^j \delta T^0_j  \ .
\ea

The energy-momentum conservation ($T^{\mu \nu}_{;\mu} = 0$) provides two equations relating the metric potentials and perturbations in each fluid. In the Newtonian gauge they are \cite{Ma:1995ey}:
\begin{eqnarray}
\label{eq:conservation}
	\dot{\delta} &=& - (1+w) \left(\theta-3{\dot{\Phi}}\right) - 3{\dot{a}\over a} \left({\delta P \over \delta\rho} - w \right)\delta \,,\nonumber\\
	\dot{\theta} &=& - {\dot{a}\over a} (1-3w)\theta - {\dot{w}\over 1+w}\theta + {\delta P/\delta\rho \over 1+w}\,k^2\delta - k^2 \sigma + k^2 \Psi \ .
\end{eqnarray}
where $w=P/\rho$. The anisotropic stress $\sigma$ vanishes for baryons and CDM, while for relativistic species, i.e. photons and neutrinos, it is generated through free-streaming and is related to $\delta$ and $\theta$ via Boltzmann equations.

Two more equations are needed to close the system of equations for metric and energy-momentum equations. For example, in GR, the Poisson equation relates the comoving density perturbations to the metric potential $\Phi$, while the anisotropy equation relates the two metric potentials and the anisotropic stress:
\ba
&&k^2\Phi = - 4 \pi G a^2 \rho \Delta \label{poisson-GR} \, , \\
&&k^2(\Phi - \Psi) = 12 \pi G a^2  (\rho + P) \sigma \label{anisotropy-GR} \, .
\ea
In an alternative theory of gravity these relations would generally be different. To parametrize possible departures from the 
$\Lambda$CDM growth at late times (\eg~redshifts $z<30$ when the contribution of relativistic species can be neglected) the following parametrization was used in the original version of MGCAMB \cite{Zhao:2008bn}:
\ba
\label{parametrization-Poisson}
&&k^2\Psi=-4 \pi G a^2\mu(k,a)\rho\Delta \,, \\
\label{parametrization-anisotropy}
&&\f{\Phi}{\Psi}=\gamma(k,a) \,,
\ea
where the two scale- and time-dependent functions $\mu(k,a)$ and $\gamma(k,a)$ were introduced to encode any modification to (\ref{poisson-GR}) and (\ref{anisotropy-GR}). Note that, in this formulation, it is the Newtonian potential $\Psi$ that appears on the right hand side of (\ref{parametrization-Poisson}), while the curvature potential $\Phi$ appears in the original Poisson equation (\ref{poisson-GR}). This is motivated by the fact that none of the observables depends solely on $\Phi$. For example, weak lensing probes the combination ($\Phi+\Psi$), while clustering of matter and the peculiar velocities are most directly related to $\Psi$, which follows from the conservation equations (\ref{eq:conservation}). Thus, measurements of peculiar velocities and galaxy counts (up to bias factors) can constrain the function $\mu$ most directly. Still, there can be circumstances in which it is advantageous to use other related parametric forms, e.g. when making a connection with particular theories \cite{Pogosian:2010tj}. We discuss some of the alternatives in \ref{alternatives}.

It should be emphasized that $\mu$ and $\gamma$ do not necessarily have a simple form in specific models of MG, and generally depend on the choice of the initial conditions\footnote{LP acknowledges very helpful discussions of this and related issues with Pedro Ferreira, Alessandra Silvestri and Constantinos Skordis} \cite{Skordis09,Ferreira:2010sz,Baker:2011jy}. For instance, in scalar-tensor models of gravity, the ratio of $\Phi$ and $\Psi$ is not a fixed function of $k$ and $a$. Instead, it is an expression that involves the time derivatives of $\Psi$ and $\Phi$. This means that $\mu$ and $\gamma$ correspond to {\it solutions} of equations of motion of a theory, rather than being a general prediction of a theory.

It is also important to make the distinction between the formalism, based on writing modified linearized Einstein equations in terms of functions $\mu$ and $\gamma$, and any particular parametric form that one may chose for these functions. The formalism is self-consistent, and satisfies the super-horizon consistency condition \cite{Wands:2000dp,Bertschinger:2006aw} as long as $(k/(aH))^2/(\mu\gamma) \rightarrow 0$ in the $k/(aH) \rightarrow 0$ limit \cite{Pogosian:2010tj}. It allows for model-independent tests of GR by simply looking for deviations of the functions from unity. If one wanted to use this formalism to compute observables in a specific theory one, strictly speaking, would have to first solve the equations of motion for perturbations in that theory, find $\mu$ and $\gamma$ by taking the appropriate ratios of the solutions, then substitute them into MGCAMB. Another approach, which would involve a degree of approximation, would be to design parametric forms for $\mu$ and $\gamma$ which can mimic solutions of particular theories \cite{Pogosian:2007sw,Bertschinger:2008zb,Giannantonio:2009gi,Thomas:2011pj}. One needs to be careful when interpreting the constraints on $\mu$ and $\gamma$ in terms of theoretical parameters, paying attention to the choice of the initial conditions and other possible assumptions and simplifications. We discuss a couple of specific parametric forms in \ref{specific}. 
 
Eqs.~(\ref{parametrization-Poisson}) and (\ref{parametrization-anisotropy}) are only applicable to late times (low redshifts) since the anisotropic stress term, which is important at epochs when radiation and relativistic neutrinos are relevant, is ignored. Here, we generalize these equations to be valid at all times:
\ba
&&k^2\Psi = - \mu(k,a) 4 \pi G a^2 \lbrace \rho \Delta + 3(\rho + P) \sigma \rbrace  \label{mg-poisson} \, , \\
&&k^2[\Phi - \gamma(k,a) \Psi] = \mu(k,a)  12 \pi G a^2   (\rho + P) \sigma \label{mg-anisotropy} \, ,
\ea
where, in Eq.~(\ref{mg-poisson}), the anisotropic $\sigma$ is added so that Eqs.~(\ref{poisson-GR}) and (\ref{anisotropy-GR}) are correctly recovered in the $\mu = \gamma=1$ limit. In this form, the equations include all the species in the universe and can be used for all times, assuming one uses appropriate initial conditions. In the current version of the MGCAMB, we kept the CAMB default initial conditions corresponding to $\gamma=1$ when a given $k$ is well outside the horizon during the radiation era ($\mu$ becomes irrelevant on superhorizon scales \cite{Pogosian:2010tj}). We plan to generalize the initial conditions to be consistent with a general choices of $\gamma$ in future upgrades of MGCAMB.   Evolving Eqs.~(\ref{mg-poisson}) and (\ref{mg-anisotropy}) with given functions $\mu$ and $\gamma$, along with the conservation and Boltzmann equations, allows us to compute various cosmological observables that can then be compared to the $\Lambda$CDM results. 

\

\subsection{Specific parametric forms of $\mu$ and $\gamma$}
\label{specific}

While MGCAMB works for any form of $\mu(k,a)$ and $\gamma(k,a)$ supplied by the user, we have coded in several specific parametrizations previously used in the literature. For example, a particular form of $\mu$ and $\gamma$ that holds approximately in some classes of $f(R)$ and scalar-tensor theories was introduced in
\cite{Bertschinger:2008zb}, hereafter referred to as the BZ parametrization:
\begin{eqnarray}
&&\mu(k,a)=\frac{1+\beta_1\lambda_1^2\,k^2a^s}{1+\lambda_1^2\,k^2a^s} \,, \nn \\
\label{BZ}
&&\gamma(k,a)=\frac{1+\beta_2\lambda_2^2\,k^2a^s}{1+\lambda_2^2\,k^2a^s} \ ,
\end{eqnarray}
where $\beta_i$'s are dimensionless couplings and $\lambda_i$'s have dimension of length. The previous version of MGCAMB \cite{Zhao:2008bn} has been used to calculate cosmological observables and to forecast errors on the five BZ parameters for future weak lensing surveys, such as DES and LSST, in combination with CMB data from Planck. 

Specific sets of BZ parameter values can be chosen to correspond to particular $f(R)$ \cite{Amendola:2006we,Song:2006ej,Hu:2007nk,Appleby:2007vb,Starobinsky:2007hu,Pogosian:2007sw} and Chameleon type \cite{Khoury:2003aq,Brax:2005ew} models. Moreover, not all of the parameters are independent. For example, $f(R)$ models with action
\begin{equation}
S = \frac{1}{16 \pi G}\int d^4x \sqrt{-g}[R+f(R)+\mathcal{L}_{\rm m}] \ ,
\label{fRaction}
\end{equation}
can be tuned to reproduce any background expansion history, and the remaining relevant quantity is the squared Compton wavelength of the new scalar degree of freedom $f_R \equiv df /dR$ mediating the fifth force. In units of the Hubble length squared it is given by \cite{Hu:2007nk,Song:2006ej}
\begin{eqnarray}
B \equiv {f_{RR} \over 1+f_R} {d R \over d\ln a} \left(  d\ln H \over d\ln a \right)^{-1}\,,
\end{eqnarray}
where $R$ is the background Ricci scalar. Thus, for a fixed background expansion history, different $f(R)$ models can be parametrized by the parameter $B_0$, which is the value of $B$ today. It was suggested in \cite{Giannantonio:2009gi} that for $B_0 \lesssim 1$ the large scale growth in $f(R)$ models can be modelled using a BZ form  with $\beta_1 = 4/3$,  $ \lambda_2^2 =  \beta_1  \lambda_1^2$ and $\beta_2 = 1/2$ as: 
\be\label{eq:par_mu}
\mu(k,a)=\frac{1}{1-1.4 \cdot 10^{-8}|\lambda_1|^2a^3}\frac{1+\frac{4}{3}\lambda_1^2\,k^2a^4}{1+\lambda_1^2\,k^2a^4}\,,
\ee
where the pre-factor is introduced to account for the background rescaling of the Newton's constant which can be important in some models, $\lambda_1^2=B_0\,c^2/(2H_0^2)$, and $\gamma$ is given by the BZ form (\ref{BZ}). We will use this parametrization in the next section to demonstrate the use of the updated MGCAMB and to constrain the $B_0$ parameter.

Similarly, models with a Yukawa--type dark matter interaction, e.g the Chameleon type models on large scales, can also be approximately expressed in terms of the extended BZ parametrization of \cite{Giannantonio:2009gi}. In this type of models, the allowed ranges of the BZ parameter values are $0 < B_0 = 2 \lambda_1^2 H_0^2/c^2 < 1$, $ 0 < \beta_1 < 2 $ and $1 <s <4$, while $\lambda_2^2 =  \beta_1  \lambda_1^2$ and $\beta_2 = 2/\beta_1 -1$, where the parameter $B_0$ is again related to the Compton wavelength of the extra scalar degree of freedom. As shown in \cite{Giannantonio:2009gi}, and in the next section, the data considered in this paper cannot constrain parameters $B_0$ and $s$ in this model.

Parametric forms such as (\ref{BZ}) or (\ref{eq:par_mu}) are certainly not guaranteed to be accurate in representing solutions of all $f(R)$, Chameleon or other MG models at all values of the theoretical parameters. An extensive investigation of the range of their applicability would be a worthwhile pursuit which, however, is outside the scope of this paper. The validity of the original BZ form (\ref{BZ}) was questioned in \cite{Thomas:2011pj}, where a more accurate parametric form specific to $f(R)$ was suggested. In Section \ref{results} we demonstrate the use of MGCAMB by working with the modified BZ form (\ref{eq:par_mu}) of \cite{Giannantonio:2009gi} that has the additional pre-factor in $\mu$. This pre-factor becomes important at $B_0 \sim 1$, which we confirmed by reproducing the CMB spectrum plots in \cite{Song:2006ej,Song:2007da} for $B_0 \lesssim 2$ where exact $f(R)$ equations of motion were used\footnote{LP thanks Yong-Seon Song for a useful discussion of this issue}. We also note that our bounds on the f(R) parameter $B_0$ are in agreement with \cite{Lombriser:2010mp}, who used the PPF parametrization of $f(R)$ solutions described in \cite{Hu:2008zd}.

\subsection{Alternative parametrizations}
\label{alternatives}

The parametrized modification of perturbed Einstein equations defined by Eqs.~(\ref{mg-poisson}) and (\ref{mg-anisotropy}) is not unique. Other forms may be useful when working with specific cosmological observables or when testing particular theories. Generally, it should be possible to express these other parametrizations in terms of $\mu$ and $\gamma$. For example, in certain circumstances, it can be desirable to work with functions $\mu$ and $\Sigma$, defined as 
\ba
\label{musigma}
&&k^2\Psi=-4 \pi G a^2\mu(k,a)\rho \tc{\Delta} \,, \nonumber \\
&&k^2(\Phi + \Psi) =8\pi G a^2\Sigma(k,a)\rho \tc{\Delta} \, .
\ea
In \cite{Song:2010fg} it was shown that working with these functions clearly captures the complementary information in weak lensing and peculiar velocity experiments. In the limit of negligible anisotropic stress $\sigma$, the function $\Sigma$ is simply related to $\mu$ and $\gamma$.  This, and other similar possibilities were discussed in \cite{Pogosian:2010tj} and it should be easy to modify MGCAMB to work in such cases.

In \cite{Bean:2010zq}, functions $Q(k,a)$ and $R(k,a)$ were introduced as 
\ba
&&k^2\Phi = - 4 \pi G a^2  Q \rho \Delta \,, \nn  \\
&&k^2(\Psi - R \Phi) = - 12 \pi G a^2 Q  (\rho + P) \sigma \ ,
\label{Bean}
\ea
which we include as an alternative parametrization in MGCAMB. Namely, when the appropriate option is selected, Eqs.~(\ref{Bean}) are evolved in place of (\ref{mg-poisson}) and (\ref{mg-anisotropy}). Functions $Q$ and $R$ are simply related to $\mu$ and $\gamma$ in the limit of negligible anisotropic stress $\sigma$:
\be 
\label{mugammatoqr}
Q = \mu \gamma  \ , \ \ R = \gamma^{-1} \ ,
\ee
although we do not rely on this conversion in practice. As a consistency check, we have reproduced some of results in \cite{Bean:2010zq} using MGCAMB.

Another commonly used parameter is the growth index $\gamma$,  proposed in Ref. \cite{Linder:2005in}. In this paper we will denote the growth index $\gamma$ as $\gamma_L$ to distinguish it from our function $\gamma(k,a)$. It is defined via
\be
f \equiv {d\ln\Delta\over d \ln a}=[ \Omega_m(a)]^{\gamma_L} \ ,
\label{def:gamma}
\ee
where $\Omega_m(a)=\rho_m/\rho_{\rm tot}$. Note, however, that this parametrization assumes that the growth function is scale-independent, so it can only be used to probe for time-dependent modifications of growth. A significant deviation of the observed $\gamma_L$ from its predicted value of $6/11 \approx 0.55$ would indicate a breakdown of $\Lambda$CDM. It is possible to use MGCAMB to constrain parameter $\gamma_L$ by expressing $\mu(k,a)$ in terms of $\gamma_L$ as \cite{Pogosian:2010tj}
\be
\mu=\frac{2}{3}\Omega_m^{\gamma_L-1}\left[\Omega_m^{\gamma_L}+2+\frac{H'}{H}+\gamma_L\frac{\Omega_m'}{\Omega_m}+\gamma_L'\ln\left(\Omega_m\right)\right] \ ,
\label{mu-gamma}
\ee
where the prime denotes differentiation with respect to $\ln a$, and the expression holds for a time-dependent $\gamma_L$.

\section{The patch}
\label{MGCAMB}

CAMB \cite{camb,Lewis:1999bs} is a publicly available code that evolves the Boltzmann and Einstein equations to calculate cosmological observables such as the CMB and matter power spectra. The first version of MGCAMB \cite{Zhao:2008bn} evolved the same equations as CAMB up to redshift $z=30$, while at lower redshifts it used Eqs.~(\ref{parametrization-Poisson}) and (\ref{parametrization-anisotropy}), which neglect effects of relativistic species. Hence, as mentioned above, it was not suitable for describing modifications at early times ($z \geq 30$). This limitation is removed in the current version of MGCAMB, which uses Eqs.~(\ref{mg-poisson}) and (\ref{mg-anisotropy}) valid for the entire redshift range and accounts for all the species.

To implement the Newtonian gauge parametrization of (\ref{mg-poisson}) and (\ref{mg-anisotropy}) in CAMB, we first need to derive an equivalent system of equations in the synchronous gauge. We have \cite{Ma:1995ey}
\ba
&&\Psi=\dot{\alpha}+\mathcal{H}\alpha  \label{psi_sync} \,,\\
&&\Phi=\eta-\mathcal{H}\alpha  \label{phi_sync} \,,
\ea
where $\alpha=(\dot{h}+6 \dot{\eta})/2k^2$ with $h$ and $\eta$ being the scalar metric potentials in the synchronous gauge defined in (\ref{sync-metric}). Our Eqs.~(\ref{mg-poisson}) and (\ref{mg-anisotropy}) then convert to 
\ba
&&k^{2}(\dot{\alpha}+\mathcal{H}\alpha)=- {\kappa \over 2}\mu(k,a) \lbrace  \rho \Delta + 3(\rho + P) \sigma \rbrace \label{mg-poisson-sync} \,, \\
&&\eta-\mathcal{H}\alpha - \gamma (\dot{\alpha}+\mathcal{H}\alpha) =  {3 \kappa \over 2k^2} \mu (\rho + P) \sigma \label{mg-anisotropy-sync}  \,,
\end{eqnarray}
where $\kappa \equiv 8 \pi G a^2$. Following the notation in CAMB, we can define the perturbation to the expansion rate, $\mathcal{Z}$, and shear, $\sigma^*$ (not to be confused with the anisotropic stress $\sigma$), as
\begin{eqnarray}
&&\mathcal{Z} = \frac{\dot{h}}{2k}  \label{z} \,, \\
&&\sigma^* = k \alpha  \label{sigma}  \,.
\label{Z-sigma}
\end{eqnarray}
In CAMB, the perturbed Einstein equations are used to evaluate these quantities:\footnote{Note that the corresponding equations in ``CAMB Notes'' are written in a different notation: their $\dot{h}$ is equivalent to $\dot{h}/6$ in our notation}
\begin{eqnarray}
&&k^2 \eta =k \mathcal{H} \mathcal{Z} - \frac{1}{2} \kappa  \rho \Delta\,, \\
&&\frac{2}{3} k^2 ( \sigma^* - \mathcal{Z})=  \kappa \rho q \, , 
\end{eqnarray}
where $\rho q = (\rho +P) v$. As we are interested in testing modifications to GR, we can no longer use Einstein's equations, and instead need an alternative way of evaluating $\mathcal{Z}$ and $\sigma^*$ based on Eqs.~(\ref{mg-poisson}) and (\ref{mg-anisotropy}) combined with conservation and Boltzmann equations. We do this by deriving $\alpha$ and $\dot{h}= 2k^2 \alpha - 6 \dot{\eta} $ from known quantities, as explained below. 

To derive $\alpha$, we substitute (\ref{mg-poisson-sync}) into (\ref{mg-anisotropy-sync}) and obtain\footnote{When evolving the equations of motion in CAMB, $\eta$ is known from the previous time step}
\begin{equation}
\label{alpha}
\alpha =  \left\lbrace \eta + \frac{\mu \kappa}{2k^2} \left[ \gamma \rho \Delta + 3(\gamma -1) (\rho + P) \sigma \right]  \right\rbrace / \mathcal{H} \ ,
\end{equation}
where $\Delta$ is in synchronous gauge. Then, to get $\dot{\eta}$, we first substitute $\dot{\alpha}$ from (\ref{mg-poisson-sync}) into (\ref{mg-anisotropy-sync}) to write
\begin{equation}
\eta = \mathcal{H}\alpha -\frac{\mu \kappa \rho}{2k^2} \Gamma \label{eta} \ ,
\end{equation}
where
\begin{equation}
\Gamma = \gamma \Delta + 3(1+w) \sigma (\gamma -1) \ . 
\end{equation}
The time derivative of (\ref{eta}) gives
\begin{equation}
\dot{\eta} =  \dot{\mathcal{H}}\alpha + \mathcal{H}\dot{\alpha} -\frac{\mu \kappa \rho}{2k^2} \lbrace 2 \mathcal{H} \Gamma - 3 \mathcal{H} (1+w) \Gamma + \frac{\dot{\mu}}{\mu} \Gamma + \dot{\Gamma} \rbrace \label{etadot} \ .
\end{equation}
Using the conservation equations in the above, we have the final expression for $\dot{\eta}$:
\begin{eqnarray}
\dot{\eta} &=&\frac{\kappa \rho }{2 \mathcal{D}} \lbrace (1+w) \left[  \mu \gamma \theta \left( 1+ \frac{3 \kappa \rho}{2k^2} (1+w) \right)+  k^2 \alpha (\mu \gamma -1) \right] + \Delta \left[ \mu(\gamma-1) \mathcal{H} -\dot{\mu} \gamma - \dot{\gamma} \mu \right]   \nonumber \\ 
 &+& 3 \dot{\sigma} (1+w) (1-\gamma) \mu  + 3 \sigma (1+w) \left[3 w \mu (\gamma -1) \mathcal{H} -(\gamma-1) \dot{\mu} -\mu \dot{\gamma} \right]  \rbrace \label{etadotfinal} \ ,
\end{eqnarray}
where $\mathcal{D}$ is 
\begin{equation}
\mathcal{D} = k^2 +\frac{3 \kappa}{2}\gamma \mu \rho ( 1+w) \ .
\end{equation}
With $\alpha$, $\dot{\eta}$ and $\dot{h}$ in hand, we can evaluate $\cal{Z}$ and $\sigma^*$ in Eqs.~(\ref{z},\ref{Z-sigma}), which are then used in the remaining equations of CAMB.

\section{A worked example: Joint constraints on massive neutrinos and modified gravity}
\label{results}

To demonstrate the use of MGCAMB in conjunction with CosmoMC, we performed a joint analysis of CMB, ISW and SNe data to constrain a particular set of modified gravity (MG) parameters together with the neutrino mass. Massive neutrinos can modify the growth of structure on different scales and redshifts depending on their mass. This modification can, in principle, be degenerate with the effects of MG on the overlapping scales and redshifts. To study the degeneracy between MG parameters and basic cosmological parameters, including the neutrino mass, we chose two specific parametrized MG models, namely, the $f(R)$ and Yukawa-type models introduced in \cite{Giannantonio:2009gi} (see Eq.~(\ref{eq:par_mu})). Cosmological constraints on $f(R)$ models were also studied in \cite{Song:2007da}.

\begin{figure}[h]
\begin{center}
\includegraphics[width=0.6\textwidth]{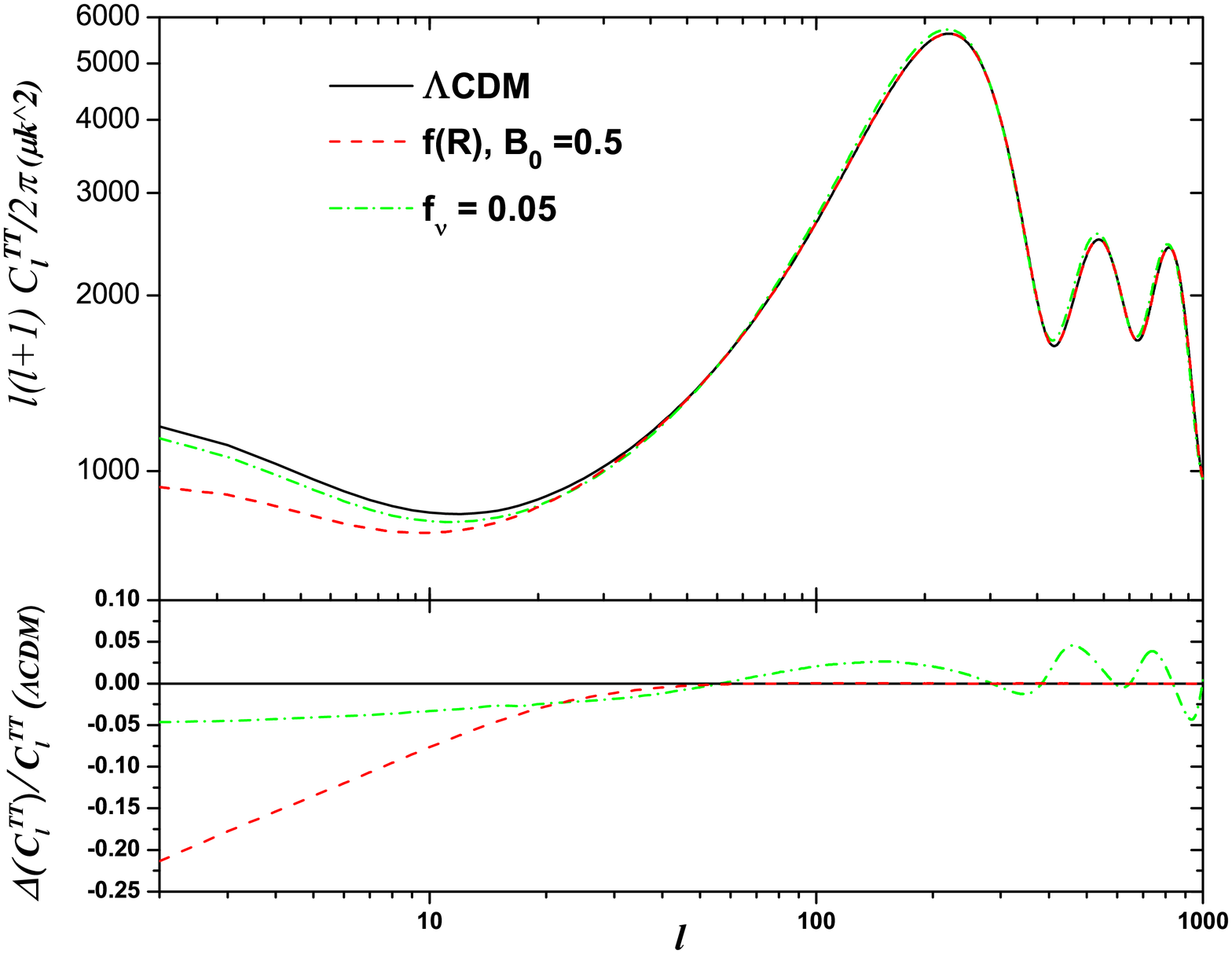}
\caption{\label{fig-MGnu}
{CMB TT power spectrum for $\Lambda$CDM (black solid), $f(R)$ model with $B_0 = 0.5$ (red dashed) and $\Lambda$CDM+massive neutrino models with $f_{\nu} = 0.05$ (green dash-dotted). }}
\end{center}
\end{figure}

On linear scales, modified gravity models, \eg~$f(R)$, generally predict enhanced growth within the Compton wavelength due to the fifth force, while massive neutrinos can damp the structure within its free-streaming scale, which is determined by the neutrino mass. Therefore, it is interesting to study the degeneracy between the MG parameters and the neutrino mass. The new version of MGCAMB released with this paper is just the tool needed to perform this analysis. Fig \ref{fig-MGnu} shows the CMB TT power spectra for $\Lambda$CDM, an "$f(R)$ model" described by Eq.~(\ref{eq:par_mu}) with $B_0 = 0.5$ and the case of massive neutrinos with $f_{\nu} \equiv \Omega_{\nu}/\Omega_{CDM} = 0.05$ corresponding to $\sum m_{\nu} \simeq 0.5 $eV. 
We attempt to constrain four models: the $f(R)$ and Yukawa-type, described by (\ref{eq:par_mu}), each with and without the presence of massive neutrinos. In all the cases, we vary the 7 primary parameters as explained in Table~\ref{tab:parameters}. In addition, we vary $f_{\nu}$ for the cases with neutrinos, $B_0$ in the case of $f(R)$ and  $\{B_0,\beta_1,s\}$ for the Yukawa model. 

We utilized MGCAMB to calculate the CMB angular spectrum and the CMB/matter cross-correlation spectrum, as well as the SNe and BAO, and used a modified version of CosmoMC to fit the models to data. We used the temperature-temperature (TT) and temperature-polarization (TE) power spectra from the WMAP seven year observation \cite{Komatsu:2010fb}, and the ISW-large scale structure (LSS) cross-correlation data by Ho et al \cite{Ho:2008bz} (see Appendix \ref{ho_implementation} for the details of the implementation). We also use the SNe data (SDSS compilation) \cite{Kessler:2009ys}, HST data from \cite{Riess:2009pu} and BAO data from \cite{Percival:2009xn} to further constrain the background expansion history. In addition, we put a top hat prior of $[10, 20]$~Giga years on the age of the universe. Given the observational data, we use CosmoMC to sample the parameter space using the Metropolis-Hastings algorithm. Table \ref{tab:parameters} shows the parameters we used for sampling and the assumed priors. We run several chains, and obtained the constraints after the chains converge perfectly.

\begin{table*}
\begin{tabular}{|l|l|c| c |}
\hline
Parameter & Explanation & \multicolumn {2}{c|}{range (min, max)}  \\
\hline
 & \multicolumn{1}{c|}{\it Primary parameters} & $f(R)$ & Yukawa--type \\
\hline
$\omega_b$ & physical baryon density; $\omega_b = h^2\Omega_b$ & \multicolumn {2}{c|}{$(0.005,
0.100)$} \\
$\omega_c$ & physical CDM density; $\omega_c = h^2\Omega_c$ &  \multicolumn {2}{c|}{$(0.01, 0.99)$}  \\
$\vartheta_*$ & sound horizon angle; $\vartheta_* = 100 \cdot r_s(z_\ast)/D_A(z_\ast)$ &
\multicolumn {2}{c|}{$(0.5, 10.0)$ }  \\
 $\tau$ & optical depth to reionisation & \multicolumn {2}{c|}{$(0.01, 0.80)$ } \\
$\ln (10^{10} A_s^2)$ & $A_s$ is the scalar primordial amplitude  &  \multicolumn {2}{c|}{$(2.7, 4.0)$} \\
$A_{SZ}$ & amplitude of the SZ template for WMAP and ACBAR & \multicolumn {2}{c|}{$(0, 2)$ } \\
$n_s$ & spectral index of primordial perturbations; $n_s - 1 = d\ln P/d\ln 
k$ &  \multicolumn {2}{c|}{$(0.5, 1.5)$} \\
\hline  
 & \multicolumn{1}{c|}{\it Neutrino parameters} & $f(R)$ & Yukawa--type \\
 \hline
$f_{\nu} $ & fraction of dark matter density as massive neutrinos  & \multicolumn {2}{c|}{$(0.0, 0.1)$} \\
\hline
 & \multicolumn{1}{c|}{\it MG parameters} & $f(R)$ & Yukawa--type \\
 \hline
$ B_0   $        &  present lengthscale of the theory (in units of the horizon scale)  &  $(0, 1)$ & $(0, 1)$ \\
$\beta_1$         &   coupling              & $ 4/3 $ & $(0.001, 2)$ \\
$s$               &   time evolution of the scalaron mass             & $ 4 $  & $(1, 4)$ \\
\hline
\end{tabular}
\caption{List of the parameters used in the Monte Carlo sampling. The ranges of the flat priors are given if a parameter is varied, or the value is given if the parameter was fixed.}
 \label{tab:parameters}
\end{table*}

\begin{figure}[h]
\begin{center}
\includegraphics[width=0.52\textwidth]{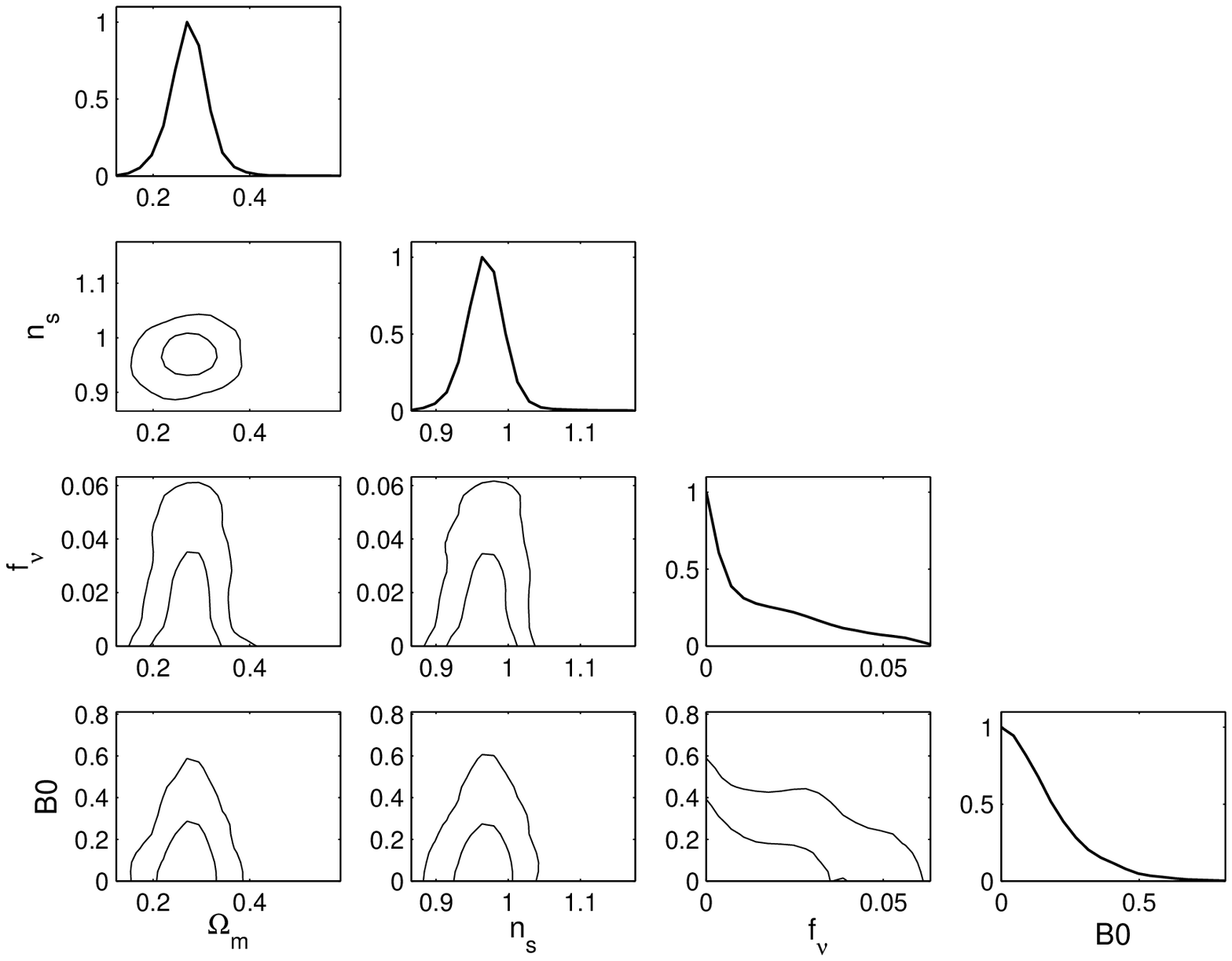}
\includegraphics[width=0.39\textwidth]{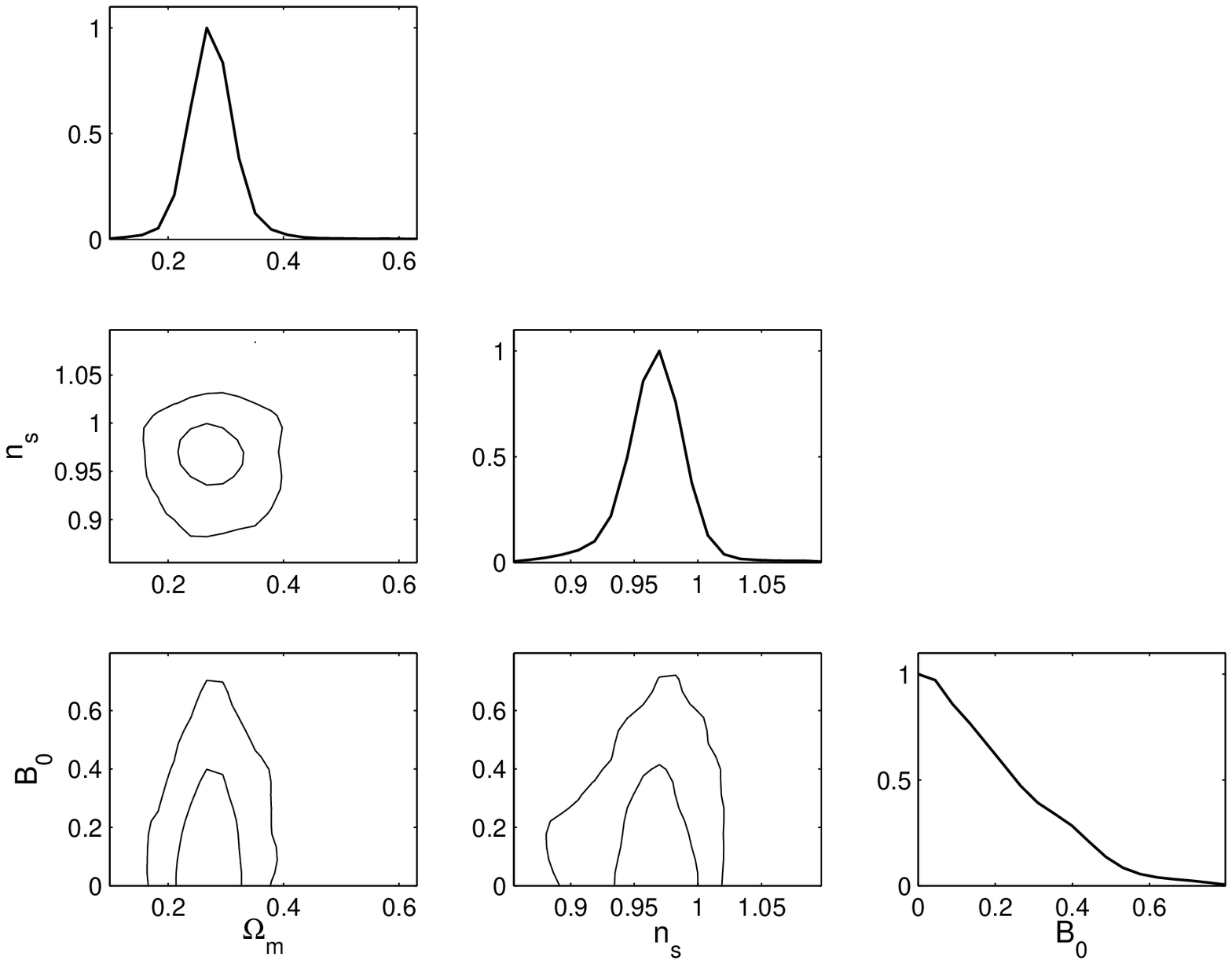}
\caption{\label{fig-fRMC}
Marginalized posterior distribution for the $f(R)$ model parameters and 2-D contour plots showing the ranges of and correlations between parameters of interest and the $68$ and $95$\% confidence limits, left: with massive neutrinos, right: without massive neutrinos.}
\end{center}
\end{figure}

\begin{figure}[h]
\begin{center}
\includegraphics[width=0.52\textwidth]{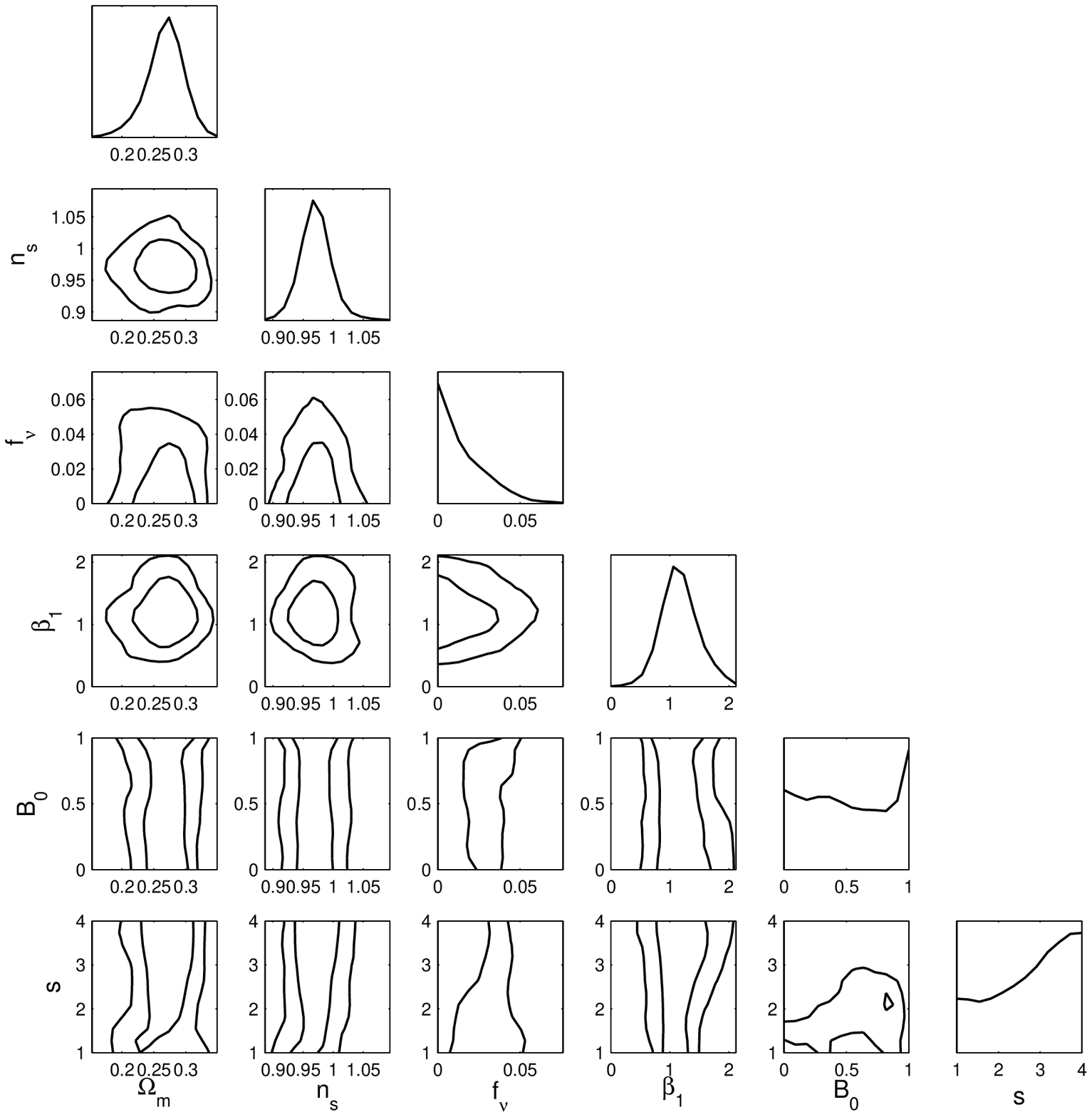}
\includegraphics[width=0.43\textwidth]{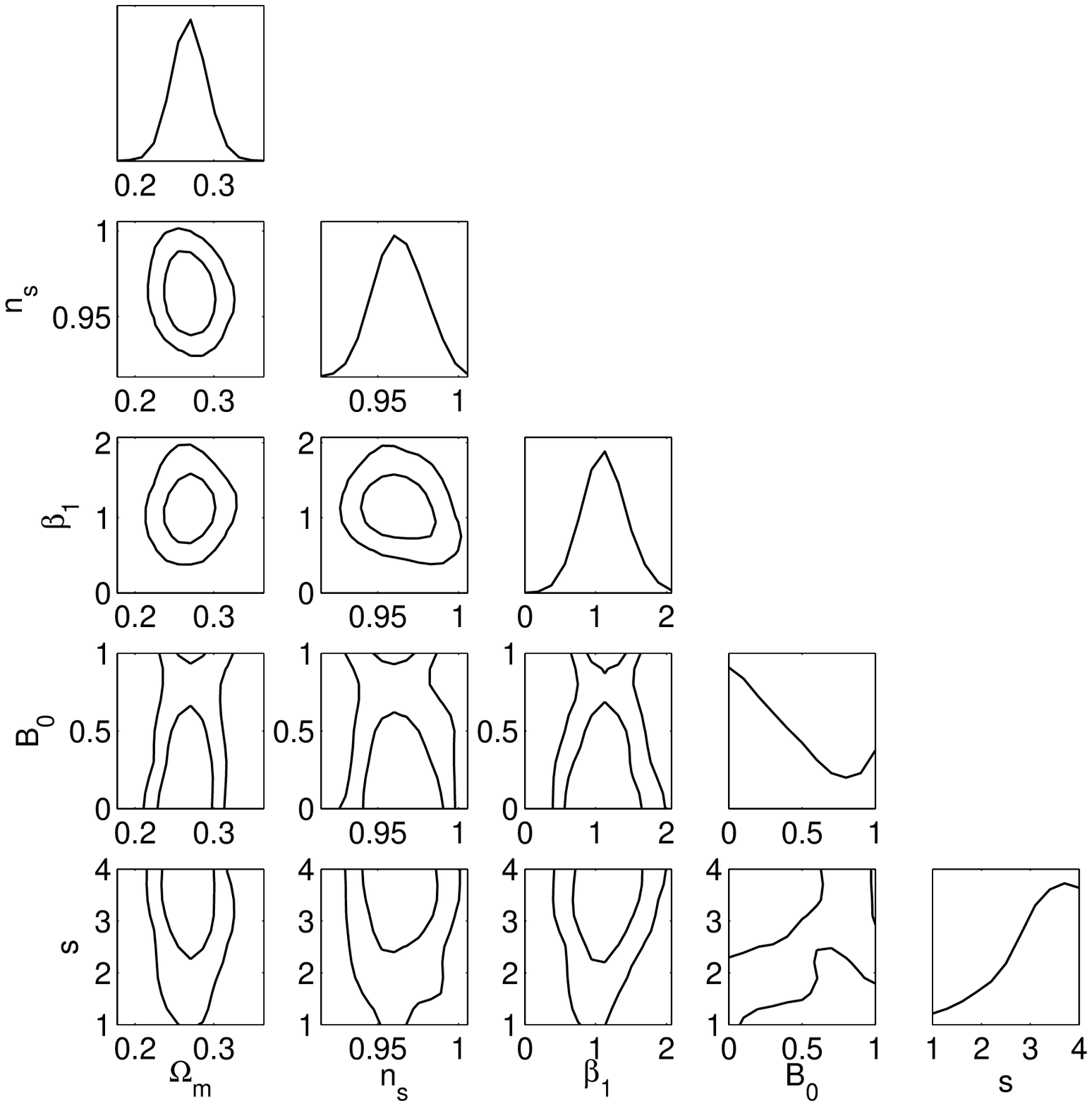}
\caption{\label{fig-yukawaMC}
Same as in Fig \ref{fig-fRMC} but for the Chameleon-type model considered in the paper.}
\end{center}
\end{figure}

Figs.~\ref{fig-fRMC} and \ref{fig-yukawaMC} show the 1-D posterior distributions, and the 2-D contour plots of the cosmological and MG parameters for the $f(R)$ and Yukawa-type models with (left panels) and without (right panels) massive neutrinos. We did not see a significant correlation between the neutrino mass and the MG parameters. This is because the MG models we have considered primarily affect the CMB spectrum via the ISW effect, which is relevant on large scales, or small $\ell$. On the other hand, the effect of (small) neutrino mass on the CMB spectrum is quite subtle and the constraint comes primarily from $\ell$ around the acoustic peaks. 

The correlation between the neutrino mass and MG would be more prominent if we considered their effect on the matter power spectrum data $P(k)$. However, adding the $P(k)$ information is non-trivial for the following reason. The MCMC module of Ho et al \cite{Ho:2008bz} is a compilation of clustering data from several surveys, and $P(k)$ from each data set is used in determining the bias, which is then used to determine the cross-correlation of clustering with CMB. So, in effect, $P(k)$ is used, but only in determining the bias, and not in constraining cosmological parameters. Same method was used in \cite{Lombriser:2010mp} and in \cite{Giannantonio:2006du}. In principle, it should be possible to include $P(k)$ into the data while properly accounting for the covariance with cross-correlation, but this task is outside the scope of this paper.

For the $f(R)$ model, we find that the $B_0$ parameter is constrained to be $B_0 < 0.4$ ($95\%$ C.L.), which corresponds to $\lambda_1^2  < 1900$ Mpc/h. The constraint on $B_0$ is practically unchanged after marginalizing over the neutrino mass. For the Yukawa model, we find that $0.7 < \beta_1 < 1.7$ ($95$\% C. L.), while $B_0$ and $s$ are very weakly constrained. In both cases, we find $f_\nu \lesssim 0.05$ at $95$\% C.L., implying $\sum m_\nu \lesssim 0.5$ eV. Our constraints on the MG parameters are consistent with those presented in Refs.~\cite{Giannantonio:2009gi,Lombriser:2010mp}. The analysis of Ref.~\cite{Giannantonio:2009gi} is based on a different compilation of ISW datasets -- we used Ho \etal~\cite{Ho:2008bz}, while they used the compilation of Giannantonio \etal~\cite{Giannantonio:2006du}. Our results are also in good agreement with those of Ref.~\cite{Lombriser:2010mp}, where the PPF framework \cite{Hu:2007pj,Hu:2008zd} employing a different set of equations was used. When the Ho \etal~\cite{Ho:2008bz} dataset was considered in \cite{Lombriser:2010mp}, it lead to a constraint of $B_0 < 0.4$ ($95\%$ C.L.) on $f(R)$ models. We hope that including the weak lensing and peculiar velocity data~\cite{Song:2010fg} and the Cosmic Mach Number (CMN) \cite{Ma:2011hf} will improve the constraints, and it is straightforward to add additional observables given the tool we developed in this work. 

\section{Summary}
\label{summary}

We have introduced a patch to CAMB and CosmoMC, called MGCAMB, that makes it possible to evaluate cosmological observables using a parametrized modification of linear Einstein equations. The motivation for this parametrization in terms of functions $\mu(a,k)$ and $\gamma(a,k)$, its consistency, and the relation to other models in the literature has been discussed in \cite{Pogosian:2010tj}. Its main value is in allowing for model-independent way of looking for departures from GR \cite{Zhao:2009fn}. However, it is also possible to derive approximate expressions for $\mu(a,k)$ and $\gamma(a,k)$ in terms of fundamental parameters of particular theories \cite{Pogosian:2007sw,Giannantonio:2009gi}.

We have demonstrated the advantages afforded by this new version of MGCAMB by obtaining joint constraints on the neutrino mass and parameters of two models of modified gravity previously considered in \cite{Giannantonio:2009gi}. We find that GR remains a good fit in all cases. In the case of $f(R)$ we find that the $B_0$ parameter is constrained to be $B_0 < 0.4$ ($95$\% C.L.),  which corresponds to $ \lambda_1^2  < 1900$ Mpc/h. In the case of Yukawa we find that  $0.7 < \beta_1 < 1.7$ ($95$\% C. L.), while $B_0$ and $s$ are very weakly constrained. In both models there is little degeneracy between the neutrino mass and the MG parameters, with the neutrino fraction constrained to be $f_\nu \lesssim 0.05$ at $95$\% C.L.. 

We have made the MGCAMB patch public at \url{http://www.sfu.ca/~aha25/MGCAMB.html} and will update it regularly to keep it compatible with the latest distributions of CAMB.

\acknowledgments

We thank Alessandra Silvestri for her input through previous and ongoing related collaborations, and Antony Lewis for helpful communications. LP acknowledges stimulating discussions with Pedro Ferreira, Alessandra Silvestri and Constantinos Skordis regarding the applicability of the parametrization used in MGCAMB, and with Yong-Seon Song regarding the CMB spectra in $f(R)$ models. AH and LP are supported by an NSERC Discovery Grant, GZ is supported by STFC grant ST/H002774/1.

\appendix
\section{Alterations to the CMB source function in CAMB.}
\label{camb_implementation}

In the notation of \cite{Zaldarriaga:1996xe}, the CMB temperature angular spectrum $C_\ell^{TT}$ can be written as
\be 
\label{spectrum}
C_\ell^{TT} = (4 \pi)^2 \int k^2 dk P(k)  [\Delta^X_\ell(k)]^2 \ ,
\ee 
where $P(k)$ is the primordial power spectrum and
\be 
\Delta^T_\ell(k) = \int_0^{\tau_0} d\tau S_T (k,\tau) j_\ell(k\tau) \ ,
\ee 
where $\tau$ is the conformal time, $S_T(k,\tau)$ is the source term, and $j_\ell(x)$ are spherical Bessel functions. In GR, the source term in terms of the synchronous gauge variables is \cite{Zaldarriaga:1996xe}
\ba 
\label{T-source}
S_T (k,\tau) &=&g\left(\Delta_{T0} +2 \dot{\alpha}+{\dot{v_b} \over k}+{\Pi \over 4 }+{3\ddot{\Pi}\over 4k^2 }\right)\nonumber \\
&+& e^{-\kappa}(\dot{\eta}+\ddot{\alpha})+\dot{g}\left(\alpha+{v_b \over k}+{3\dot{\Pi}\over {\color{red}{2}}k^2 }\right)+{3 \ddot{g}\Pi \over 4k^2} \nonumber \\
\ea
where $\kappa$ is the optical depth, $g$ is the visibility function, $\Pi = \Delta^T_2+\Delta^P_2+\Delta^P_0$ and $\Delta^T_\ell (\Delta^P_\ell)$ are the $\ell$'th moments of $\Delta^T(\Delta^P)$ in term of Legendre polynomials \cite{Zaldarriaga:1996xe}. 

In the default version of CAMB, the perturbed Einstein equations are used in the evaluation of the source $S_T(k,\tau)$. Since we have introduced modifications to the Poisson (\ref{mg-poisson-sync}) and anisotropy (\ref{mg-anisotropy-sync}) equations, it was important to avoid assuming GR in the expression for the source. The MGCAMB patch replaces the default CMB source in CAMB with the one given by Eq.~(\ref{T-source}). In addition, the evaluation of the ISW term, $e^{-\kappa}(\dot{\eta}+\ddot{\alpha})$, in (\ref{T-source}) is modified. Namely, it was originally given by
\be 
\dot{\eta}+\ddot{\alpha} =  \frac{\kappa}{2k^2} \left[ 2 (\dot{\rho} \Delta + \rho \dot{\Delta}) + \frac{3}{2}(\rho +P)\dot{\sigma} + \frac{3}{2}
(\dot{\rho}+\dot{P}) \sigma \right]\ ,
\ee
while the MGCAMB patch changes it to 
\be 
\dot{\eta}+\ddot{\alpha} = \frac{\kappa}{2k^2} \left\lbrace  - \left[(\gamma+1) (\dot{\rho} \Delta + \rho \dot{\Delta}) + \gamma \frac{3}{2}(\rho +P)\dot{\sigma} + \gamma \frac{3}{2} (\dot{\rho}+\dot{P})
 \sigma \right]  + \dot{\gamma} \mu \left[ {(\rho \Delta)} + \frac{3}{2}(\rho +P)\sigma \right] \right\rbrace  \ ,
\ee
where Eqs.~(\ref{alpha}-\ref{etadot}) are used. It can be easily checked that for $\mu = \gamma=1$, we get back the ISW term in GR.

\section{Alterations to the CMB-LSS cross-correlation patch of Ho et~al.}
\label{ho_implementation}

To use the ISW data of Ho et al.~\cite{Ho:2008bz}, we need to calculate the theoretical CMB-LSS cross-correlation angular spectrum, $C_\ell^{gT}$. Below, we explain how it is done with MGCAMB.

In \cite{Ho:2008bz},  $C_\ell^{gT}$ is written as\footnote{Note that the different pre-factor, $(2/\pi)$ in Eq.~(\ref{ISWspectrum}) rather than $(4\pi)^2$ in (Eq.~(\ref{spectrum})), is due to different Fourier conventions \cite{Hu:1997hp}.}
\begin{equation}
\label{ISWspectrum}
C^{gT}_\ell=\frac{2}{\pi} \int k^2 dk P(k) g_\ell (k) T_\ell(k) \ ,
\end{equation}
with 
\begin{equation}
g_\ell(k)=\int dz \, b_i(z) \Pi(z) D(z)j_\ell(k\chi(z)) \ ,
\end{equation}
and  
\begin{equation}
T_\ell(k)=3\frac{H_0^2}{c^2}\Omega_m T_{\rm CMB} \times\int dz \frac{d}{dz}\left[D(z)(1+z)\right]\frac{j_\ell(k\chi(z))}{k^2} \ ,
\end{equation}
where $\Pi(z)$ is the normalized selection function, $T_{\rm CMB}$ is temperature of CMB today, $\chi(z)$ is the comoving distance to redshift $z$, $b_i(z)$ is the bias factor, integration over $\tau$ is replaced with integration over redshift and the growth factor $D(z)$ is defined as 
\begin{equation}
\label{growth}
\frac{\delta(k, z)}{\delta(k,0)} = \frac{D(z)}{D(0)} \ .
\end{equation}
Expression (\ref{ISWspectrum}) can be simplified for $\ell \gtrsim 10$ by working in the flat-sky approximation and substituting $k=(\ell+1/2)/\chi(z)$, which gives \cite{Ho:2008bz}
\begin{eqnarray}
\label{clgt}
C_\ell^{gT} &=& {3\Omega_m H_0^2 T_{\rm CMB}\over c^2 (\ell+1/2)^2}  \times \int dz b(z) \Pi(z) {H(z) \over c} D(z){d\over dz}[D(z)(1+z)] P\left(\frac{\ell+1/2}\chi\right) \ .
\end{eqnarray}

In a general theory of gravity, the growth factor can have a complicated form depending both on scale and time. With Eqs.~(\ref{mg-poisson-sync}) and (\ref{mg-anisotropy-sync}), it is straight-forward to show that at late times, when the anisotropic stress due to relativistic species is negligible, Eq.~(\ref{clgt}) is modified to 
\begin{eqnarray}
C_\ell^{gT} &=& {3\Omega_m H_0^2 T_{\rm CMB}\over c^2 (\ell+1/2)^2}  \times \int dz b(z) \Pi(z) {H(z) \over c} D(z){d\over dz}[D(z)(1+z) \mu(k,z) (1+\gamma(k,z))]  P\left(\frac{\ell+1/2}\chi\right) \ ,
\end{eqnarray}
where  $D(z)$ is still defined as Eq.~(\ref{growth}) and we use MGCAMB to calculate $D(z)$ from Eq.~(\ref{growth}) at each redshift and also the modified matter power spectrum today, $ P((\ell+1/2)/\chi)$. The derivative of the growth factor with respect to redshift can be calculated using Eqs.~(\ref{eq:conservation}), (\ref{mg-poisson-sync}) and the corresponding expressions for $\mu(k,z)$ and $\gamma(k,z)$.


\begin{thebibliography}{99}

\bibitem{Linder:2007hg}
  E.~V.~Linder and R.~N.~Cahn,
  Astropart.\ Phys.\  {\bf 28}, 481 (2007)
  [arXiv:astro-ph/0701317].

\bibitem{Caldwell:2007cw}
  R.~Caldwell, A.~Cooray and A.~Melchiorri,
  Phys.\ Rev.\  D {\bf 76}, 023507 (2007)
  [arXiv:astro-ph/0703375].

\bibitem{Hu:2007pj}
  W.~Hu and I.~Sawicki,
  Phys.\ Rev.\  D {\bf 76}, 104043 (2007)
  [arXiv:0708.1190 [astro-ph]].

\bibitem{Bertschinger:2008zb}
  E.~Bertschinger and P.~Zukin,
  Phys.\ Rev.\  D {\bf 78}, 024015 (2008)

\bibitem{Fang:2008sn}
  W.~Fang, W.~Hu and A.~Lewis,
  Phys.\ Rev.\  D {\bf 78}, 087303 (2008)
  [arXiv:0808.3125 [astro-ph]].

\bibitem{Zhao:2008bn}
  G.~B.~Zhao, L.~Pogosian, A.~Silvestri and J.~Zylberberg,
  Phys.\ Rev.\  D {\bf 79}, 083513 (2009)
  [arXiv:0809.3791 [astro-ph]].

\bibitem{Giannantonio:2009gi}
  T.~Giannantonio, M.~Martinelli, A.~Silvestri and A.~Melchiorri,
  JCAP {\bf 1004}, 030 (2010)
  [arXiv:0909.2045 [astro-ph.CO]].

\bibitem{Zhao:2009fn}
  G.~B.~Zhao, L.~Pogosian, A.~Silvestri and J.~Zylberberg,
  Phys.\ Rev.\ Lett.\  {\bf 103}, 241301 (2009)

\bibitem{Daniel:2009kr}
  S.~F.~Daniel {\it et al.},
  Phys.\ Rev.\  D {\bf 80}, 023532 (2009)
  [arXiv:0901.0919 [astro-ph.CO]].

\bibitem{Daniel:2010ky}
  S.~F.~Daniel, E.~V.~Linder, T.~L.~Smith, R.~R.~Caldwell, A.~Cooray, A.~Leauthaud and L.~Lombriser,
  Phys.\ Rev.\  D {\bf 81}, 123508 (2010)
  [arXiv:1002.1962 [astro-ph.CO]].

\bibitem{Song:2010rm}
  Y.~S.~Song, L.~Hollenstein, G.~Caldera-Cabral and K.~Koyama,
  arXiv:1001.0969 [astro-ph.CO].
 
\bibitem{Bean:2010zq}
  R.~Bean and M.~Tangmatitham,
  Phys.\ Rev.\  D {\bf 81}, 083534 (2010)
  [arXiv:1002.4197 [astro-ph.CO]].

\bibitem{Zhao:2010dz}
  G.~B.~Zhao {\it et al.},
  Phys.\ Rev.\  D {\bf 81}, 103510 (2010)
  [arXiv:1003.0001 [astro-ph.CO]].

\bibitem{Song:2010fg}
  Y.~S.~Song, G.~B.~Zhao, D.~Bacon, K.~Koyama, R.~C.~Nichol and L.~Pogosian,
  arXiv:1011.2106 [astro-ph.CO].
  
\bibitem{Lombriser:2010mp}
  L.~Lombriser, A.~Slosar, U.~Seljak and W.~Hu,
  arXiv:1003.3009 [astro-ph.CO].
  
\bibitem{Dossett:2011zp}
  J.~Dossett, J.~Moldenhauer and M.~Ishak,
  arXiv:1103.1195 [astro-ph.CO].
   
\bibitem{Silvestri:2009hh}
  A.~Silvestri and M.~Trodden,
  Rept.\ Prog.\ Phys.\  {\bf 72}, 096901 (2009)
  [arXiv:0904.0024 [astro-ph.CO]].

\bibitem{Clifton:2011jh}
  T.~Clifton, P.~G.~Ferreira, A.~Padilla and C.~Skordis,
  arXiv:1106.2476 [astro-ph.CO].
   
\bibitem{camb} \url{http://camb.info/}

\bibitem{Lewis:1999bs}
  A.~Lewis, A.~Challinor and A.~Lasenby,
  Astrophys.\ J.\  {\bf 538}, 473 (2000)
  [arXiv:astro-ph/9911177].
  
\bibitem{Wands:2000dp}
  D.~Wands, K.~A.~Malik, D.~H.~Lyth and A.~R.~Liddle,
  Phys.\ Rev.\  D {\bf 62}, 043527 (2000)

\bibitem{Bertschinger:2006aw}
  E.~Bertschinger,
  Astrophys.\ J.\  {\bf 648}, 797 (2006)
  
\bibitem{MGCAMB}
\url{Http://www.sfu.ca/~aha25/MGCAMB.html}  

\bibitem{cambsources}
\url{http://camb.info/sources/}

\bibitem{Ho:2008bz}
  S.~Ho, C.~Hirata, N.~Padmanabhan, U.~Seljak and N.~Bahcall,
  Phys.\ Rev.\  D {\bf 78}, 043519 (2008)
  [arXiv:0801.0642 [astro-ph]].

\bibitem{Ma:1995ey}
  C.~P.~Ma and E.~Bertschinger,
  Astrophys.\ J.\  {\bf 455}, 7 (1995)
  [arXiv:astro-ph/9506072].

\bibitem{Skordis09}
C.~Skordis,   Phys.\ Rev.\  D {\bf 79}, 123527 (2009).

\bibitem{Ferreira:2010sz}
  P.~G.~Ferreira and C.~Skordis,
  Phys.\ Rev.\  D {\bf 81}, 104020 (2010)
  [arXiv:1003.4231 [astro-ph.CO]].

\bibitem{Baker:2011jy}
  T.~Baker, P.~G.~Ferreira, C.~Skordis and J.~Zuntz,
  arXiv:1107.0491 [astro-ph.CO].
  
\bibitem{Pogosian:2010tj}
  L.~Pogosian, A.~Silvestri, K.~Koyama and G.~B.~Zhao,
  Phys.\ Rev.\  D {\bf 81}, 104023 (2010)
  [arXiv:1002.2382 [astro-ph.CO]].
  
\bibitem{Pogosian:2007sw}
  L.~Pogosian and A.~Silvestri,
  Phys.\ Rev.\  D {\bf 77}, 023503 (2008)
  [arXiv:0709.0296 [astro-ph]].
 
\bibitem{Thomas:2011pj}
  S.~A.~Thomas, S.~A.~Appleby and J.~Weller,
  JCAP {\bf 1103}, 036 (2011)
  [arXiv:1101.0295 [astro-ph.CO]].
  
\bibitem{Amendola:2006we}
  L.~Amendola, R.~Gannouji, D.~Polarski and S.~Tsujikawa,
  Phys.\ Rev.\  D {\bf 75}, 083504 (2007)
  [arXiv:gr-qc/0612180].

\bibitem{Hu:2007nk}
  W.~Hu and I.~Sawicki,
  Phys.\ Rev.\  D {\bf 76}, 064004 (2007)
  [arXiv:0705.1158 [astro-ph]].

\bibitem{Appleby:2007vb}
  S.~A.~Appleby and R.~A.~Battye,
  Phys.\ Lett.\  B {\bf 654}, 7 (2007)
  [arXiv:0705.3199 [astro-ph]].

\bibitem{Starobinsky:2007hu}
  A.~A.~Starobinsky,
  JETP Lett.\  {\bf 86}, 157 (2007)
  [arXiv:0706.2041 [astro-ph]].
  
\bibitem{Song:2006ej}
  Y.~S.~Song, W.~Hu and I.~Sawicki,
  Phys.\ Rev.\  D {\bf 75}, 044004 (2007)
  [arXiv:astro-ph/0610532].

\bibitem{Khoury:2003aq}
  J.~Khoury and A.~Weltman,
  Phys.\ Rev.\ Lett.\  {\bf 93}, 171104 (2004)
  [arXiv:astro-ph/0309300].

\bibitem{Brax:2005ew}
  P.~Brax, C.~van de Bruck, A.~C.~Davis and A.~M.~Green,
  Phys.\ Lett.\  B {\bf 633}, 441 (2006)
  [arXiv:astro-ph/0509878].

\bibitem{Hu:2008zd}
  W.~Hu,
  Phys.\ Rev.\  D {\bf 77}, 103524 (2008)
  [arXiv:0801.2433 [astro-ph]].

\bibitem{Linder:2005in}
  E.~V.~Linder,
  Phys.\ Rev.\  D {\bf 72} (2005) 043529
  [arXiv:astro-ph/0507263].
  
\bibitem{Song:2007da}
  Y.~S.~Song, H.~Peiris and W.~Hu,
  Phys.\ Rev.\  D {\bf 76}, 063517 (2007)
  [arXiv:0706.2399 [astro-ph]].

\bibitem{Komatsu:2010fb}
  E.~Komatsu {\it et al.}  [WMAP Collaboration],
  Astrophys.\ J.\ Suppl.\  {\bf 192}, 18 (2011)
  [arXiv:1001.4538 [astro-ph.CO]].

\bibitem{Kessler:2009ys}
  R.~Kessler {\it et al.},
  Astrophys.\ J.\ Suppl.\  {\bf 185}, 32 (2009)
  [arXiv:0908.4274 [astro-ph.CO]].
  
\bibitem{Riess:2009pu}
  A.~G.~Riess {\it et al.},
  Astrophys.\ J.\  {\bf 699}, 539 (2009)
  [arXiv:0905.0695 [astro-ph.CO]].

\bibitem{Percival:2009xn}
  B.~A.~Reid {\it et al.}  [SDSS Collaboration],
  Mon.\ Not.\ Roy.\ Astron.\ Soc.\  {\bf 401}, 2148 (2010)
  [arXiv:0907.1660 [astro-ph.CO]].

\bibitem{Giannantonio:2006du}
  T.~Giannantonio {\it et al.},
  Phys.\ Rev.\  D {\bf 74} (2006) 063520
  [arXiv:astro-ph/0607572].

\bibitem{Ma:2011hf}
  Y.~Z.~Ma, J.~P.~Ostriker and G.~B.~Zhao,
  arXiv:1106.3327 [astro-ph.CO].

\bibitem{Zaldarriaga:1996xe}
  M.~Zaldarriaga and U.~Seljak,
  Phys.\ Rev.\  D {\bf 55}, 1830 (1997)
  [arXiv:astro-ph/9609170].

\bibitem{Hu:1997hp}
  W.~Hu and M.~J.~White,
  Phys.\ Rev.\  D {\bf 56}, 596 (1997)
 [arXiv:astro-ph/9702170].
  
\end{thebibliography}
\end{document}